\documentclass[prb,twocolumn]{revtex4-1} 

\usepackage{amsmath}  
\usepackage{amsfonts} 
\usepackage{graphicx} 

\begin{document}

\title{``The Physics of Life,'' an undergraduate general education biophysics course}

\author{Raghuveer Parthasarathy}
\email{raghu@uoregon.edu} 
\affiliation{Department of Physics, University of Oregon, Eugene, OR, USA, 97403-1274}

\date{\today}

\begin{abstract}
Improving the scientific literacy of non-scientists is an important goal, both because of the ever-increasing impact of science and technology on our lives, and because understanding science enriches our experience of the natural world. One route to improving scientific literacy is via general education undergraduate courses --- i.e. courses intended for students not majoring in the sciences or engineering --- which in many cases provide these students' last formal exposure to science. I describe here a course on biophysics for non-science-major undergraduates recently developed at the University of Oregon (Eugene, OR, USA). Biophysics, I claim, is a particularly useful vehicle for addressing scientific literacy.  It involves important and general scientific concepts, demonstrates connections between basic science and tangible, familiar phenomena related to health and disease, and illustrates that scientific insights develop by applying tools and perspectives from disparate fields in creative ways.  In addition, biophysics highlights the far-reaching impact of physics research.  I describe the general design of this course, which spans both macroscopic and microscopic topics, and the specific content of a few of its modules. I also describe evidence-based pedagogical approaches adopted in teaching the course, and aspects of its enrollment and evaluation.  
\end{abstract}

\maketitle

\section{Introduction} 

Few would disagree that fostering scientific literacy among the general public is a worthwhile goal. We live in a world of increasing technological complexity, and developments in biotechnology, energy use, communications, and many other fields impact people's lives in unprecedented ways. Moreover, the advance of science has illuminated countless fascinating aspects of the inner workings of nature, from the structure of stars to the interactions of genes, and an understanding of science opens the doors to an enriching understanding of these insights.

	There is widespread concern, however, that the level of scientific literacy in contemporary society is poor, with respect to both basic scientific knowledge and, more importantly, understanding of the nature of the scientific process \cite{pew2009, miller2004, gross2006}.  One way to address this is via general education undergraduate courses --- i.e. courses intended for students not majoring in the sciences or engineering --- which in many cases provide these students' last formal exposure to science. A variety of such courses exist in the Physics departments at many universities, structured as overviews of wide swathes of the subject, or forays into more specialized niches. I describe here a course on  \textit{biophysics} for non-science-major undergraduates, titled ``The Physics of Life,'' that I have recently developed and taught at the University of Oregon. 

	Biophysics, I claim, is a particularly useful vehicle for addressing scientific literacy. It involves important and general scientific concepts, demonstrates connections between basic science and tangible phenomena related to health and physiology, and illustrates how scientific insights do not develop along predictable paths, but rather often arise by the creative application of perspectives and tools from disparate fields.  Moreover, it highlights the importance of physics in biological research, a view increasingly realized among biologists \cite{ascb2012, visionandchange, hilborn2014}, but not by the general public.
 
Here I describe the design of this one quarter (ten week) course, the specific content of a few of its modules, its use of active learning and evidence-based pedagogy, and aspects of its enrollment and evaluation. The aim of the article, in addition to documenting aspects of this course, is to hopefully help seed similar classes elsewhere, or instances in which biophysical concepts are incorporated into other general education classes.

\section{Goals and Topics}

The course has three overarching goals: (1) To help students learn how physical principles guide and constrain life. This includes developing a basic understanding of what the biomolecules and biomaterials that make up organisms are, and how their physical properties and interactions govern their function. (2) To improve students' ability to understand quantitative data and models. This goal spans skills such as numerical estimation and grasping the meaning of graphs, including non-standard graphs such as log-log plots (e.g. for biomechanical scaling relationships). (3) To improve students' comprehension of the process by which scientific understanding develops. This encompasses examples of the complex relationships between ``pure'' and ``applied'' science, and of the connections between seemingly disparate fields of scientific study. Particular topics were chosen to contribute to these goals and develop students' scientific literacy. The course is part of the University of Oregon's Science Literacy Program \cite{slp}, which aims to help implement evidence-based pedagogical methods across a range of general education courses spanning several science departments, and to facilitate new classes and new approaches to faculty and student training. 

\subsection {Macroscopic topics}

The ten week term is roughly divided into two halves. The first covers macroscopic topics, with a focus on scaling concepts, i.e. understanding how various physical forces and biomechanical properties scale with organism size, and how this influences the behavior and physiology of animals and plants. 

\subsubsection{Surface tension}

\begin{figure}[h!]

\centering
\includegraphics[width=3.5in]{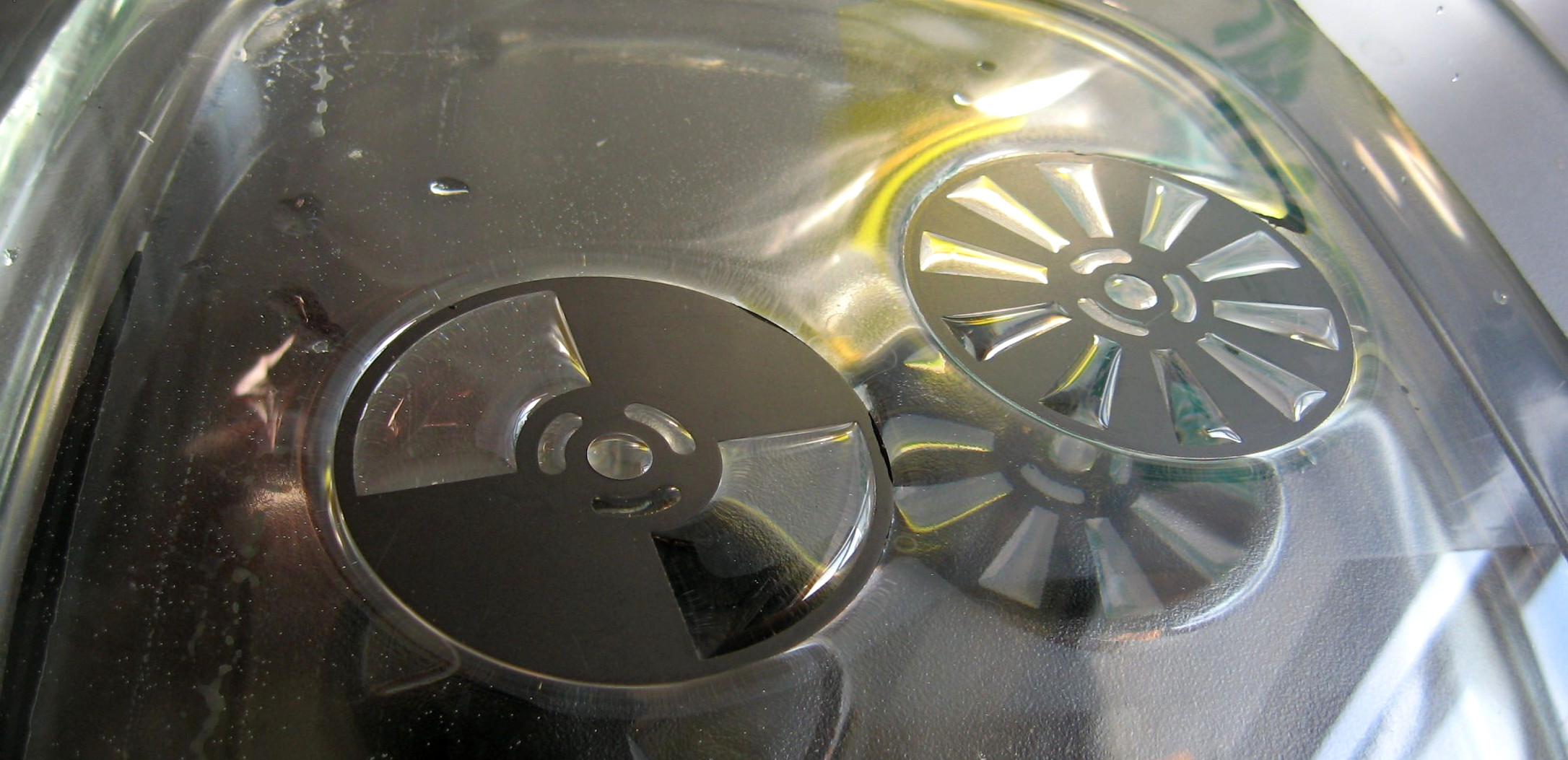}
\caption{Two chopper wheels, with the same area but different contact perimeters, are supported atop water by surface tension. Adding weights to each one until it sinks helps demonstrate that the force associated with surface tension is dependent on perimeter, which leads to an understanding of the scaling properties of interfacial forces.}
\label{chopperwheels}
\end{figure}

The first biological question we address is why small insects can walk on water, while humans cannot. This leads to the concept of surface tension, introduced by demonstrating a metal paper clip sitting atop a water surface. The simple question, ``Why does it stay up?'' is a surprisingly difficult one to answer; many students will state ``surface tension,'' but when probed will not be able to explain what this means, leading to interesting conversations on the inadequacy of simply naming phenomena as compared to understanding the mechanisms underlying them \cite{feynman}. We then discuss the nature of liquids, and how a consequence of intermolecular attraction is a tendency to minimize surface area, and hence surface tension. There is, of course, a force associated with surface tension, holding up our paper clip against the force of gravity. What geometric properties of the object should this force depend on? With two objects of equal area, but different edge lengths atop a bath of water (Figure~\ref{chopperwheels}), adding weights to each until they sink, one can show quite simply that the shape with the greater perimeter can support considerably more weight, and so has a larger surface tension force associated with it. (It's a surprisingly dramatic demonstration; the students make guesses beforehand and are almost breathless throughout the slow addition of weights.) From this, we establish that the force associated with surface tension scales with length and that, all things being equal, an organism whose dimensions double would, at a liquid surface, experience an upward force that is twice as large. We also learn that the force of gravity is proportional to the mass of an object, and hence scales as the cube of length, which points to the answer to our original question: if we imagine organisms growing in size, the force of gravity increases much more than the force that surface tension can provide. This difference in scaling behavior underlies differences in animal behavior, and explains why we don't find large animals walking on water. It also explains why an individual fire ant, for example, can walk on water due to surface tension but a raft of ants cannot (which students are able to predict, based on their improved physical understanding), which leads ants agglomerating in flooding jungles to trap air bubbles to harness the force of buoyancy to keep themselves up, the subject of recent, fascinating studies \cite{mlot_fireants}. 

This leads simply to a topic of considerable physiological importance: breathing. The surface of the lungs is wet, and so much of the work necessary for breathing is work done against surface tension. (For this reason it is easier to inflate lungs with water than with air \cite{clements62}, which students are surprised to learn.) Our lungs secrete, therefore, a surfactant that lowers the surface tension of the lungs and facilitates breathing. This surfactant is produced rather late in gestation, however, around week 30, and its absence leads to Infant Respiratory Distress Syndrome (IRDS), the leading cause of death among premature infants. The mortality rate from IRDS has dropped from about 25,000 deaths per year in the 1960s to less than 1000 in 2005 \cite{Schraufnagel}, due to the development of surfactant treatments --- in essence, injecting animal-based or synthetic amphiphilic molecules into the lungs. The connections between a basic physical concept, a biological function, and a real-world application are rarely clearer than this.

Of course, the discussion of surface tension above will seem trivial to most readers of this journal. It serves a useful function in the course, however, beyond being interesting, and that is to introduce scaling concepts. This takes a considerable amount of work --- a statement as simple as ``volume is proportional to length cubed'' is not only foreign to most students but is remarkably non-intuitive. In general, their own prior exposure to geometry has been centered on memorizing formulas for, for example, the volumes of various shapes, rather than developing more expansive notions of concepts like volume and area. I describe later in this article various exercises involving, for example, making log-log plots of area and length for simple shapes, and measuring volumes of complex shapes, that build intuition about geometric scaling relationships.

\subsubsection{Biomechanics and scaling}

\begin{figure}[h!]

\centering
\includegraphics[width=3.5in]{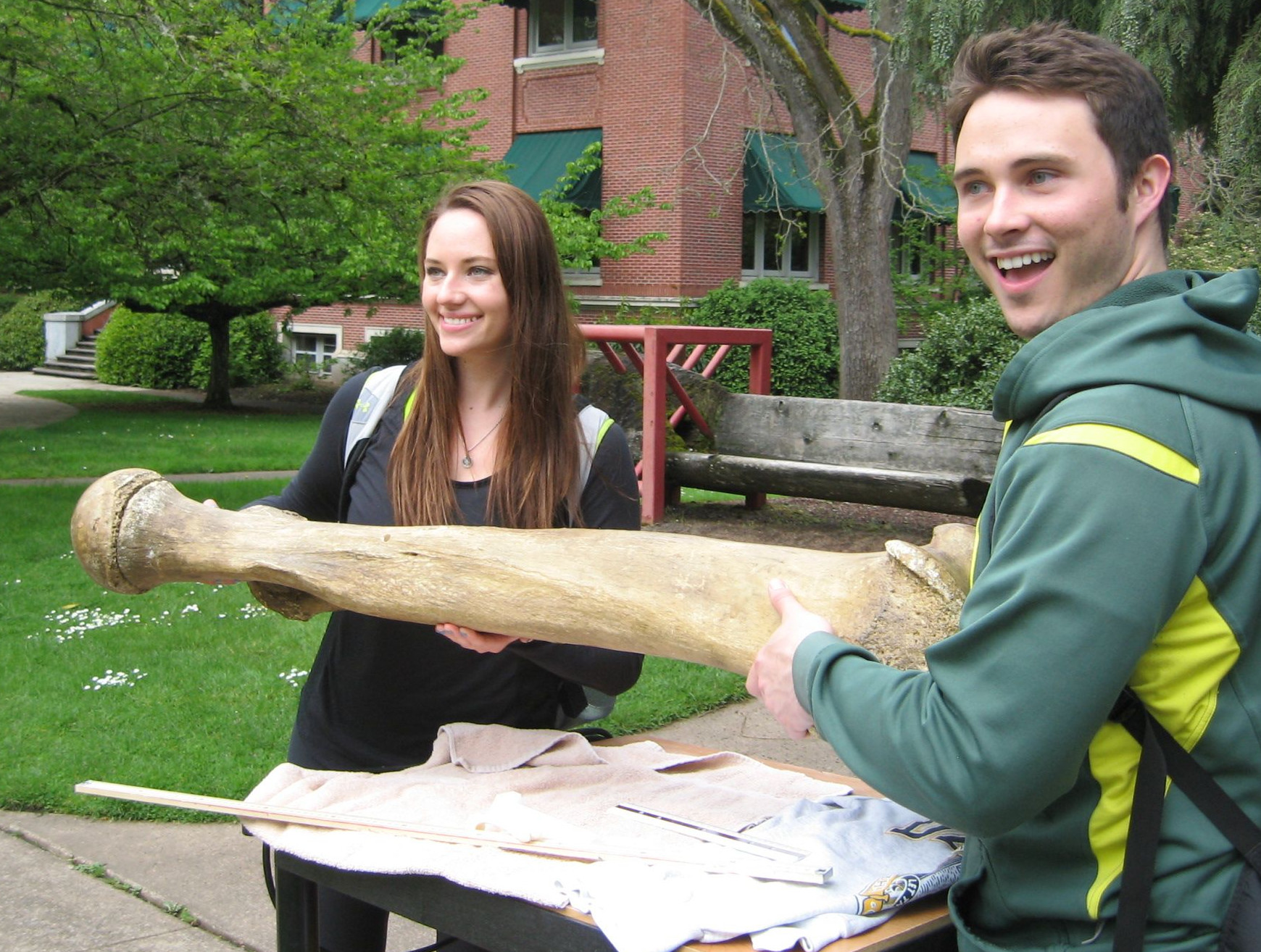}
\caption{An elephant femur helps illustrate the biophysics of bone shape, and attracts attention while being carted through campus.}
\label{elephantfemur}
\end{figure}

We then examine other issues of biomechanical scaling, especially the question of why larger land animals need disproportionately thicker bones than smaller ones. An elephant's femur, for example, is about 10 times longer than a small dog's, but has a diameter about 20 times greater. It is straightforward to illustrate this with images of bones; alternatively, one can find real bones (Figure~\ref{elephantfemur})\cite{UOelephant}. Why are the bones so disproportionate?  Again, the scaling of different physical properties provides an explanation. The force of gravity is proportional to mass, and hence length cubed, while the strength of bones, or beams in general, is proportional to their cross-sectional area and hence the square of length. (One could of course further explore the continuum mechanics of buckling and make the preceding statement more accurate; we do not in this course.) To avoid being crushed, larger animals need disproportionately wider bones.  Notably, if the bone diameter scales as length$^{3/2}$, gravitational force and bone strength follow one another; this is the case for ``mechanically similar'' animals. Following an example from McMahon and Bonner \cite{mcmahon1983}, we examine plots of bone dimensions for a wide range of bovids (antelope, wildebeest, etc.), and find that this mechanical similarity holds, highlighting a non-obvious shared characteristic of these diverse animals. The topic of bone shape and scaling has a long history, dating at least to Galileo \cite{galileo}, and is compellingly discussed in a variety of books (e.g. Refs.~\onlinecite{mcmahon1983, lifesdevices}), discussed further below.

In addition to their relevance to animal form, the biomechanical scaling ideas explored above relate to contemporary issues of human health in intriguing ways. The Body Mass Index (BMI), for example, postulates a person's mass ($M$) divided by height squared ($h^2$) as a convenient and size-invariant measure of obesity. If people of different heights were the same shape (which is not the case), one would expect $m/h^3$ to be a useful measure. The BMI assumes a particular non-isometric form, with $m \sim h^p$ and $p=2$. Do actual data on masses and heights obey this? Strikingly, they do not \cite{korevaar}; to the extent that there is power-law scaling at all, $p$ is roughly 2.6-2.7. The peculiarities of the BMI, such as the inaccuracy of its obesity implications for tall people, are familiar to many students, which leads to interesting discussions of why the measure exists and persists.

\subsection {Microscopic topics}

	The second half of the course covers microscopic topics, especially the cellular and subcellular phenomena that are the targets of most contemporary biophysical study. A key goal is to convey an understanding of random, Brownian motion and its importance. Students are used to seeing cartoons or diagrams of biological processes, and from these form the mistaken impressions that these processes are more ordered and deterministic than they actually are, and that small-scale ``machines'' can be thought of simply as scaled-down versions of macroscopic devices. Reality is, of course, much different, and seeing how dissimilar from our familiar experience the microscopic world is enhances our appreciation of it. 

\subsubsection{Brownian Motion}

We begin with observations, either with tabletop microscopes or previously recorded videos, of colloidal Brownian motion, noting also its scientific history \cite{mazo}. We introduce the idea of a random walk. To characterize this, rather than constructing an algebraic derivation, we turn again to now-familiar tools for uncovering scaling behavior, plotting various properties of random walks simulated in class and finding, eventually, that the root mean square distance traveled robustly scales as time$^{0.5}$. This non-linear scaling of distance and time, together with a few numbers, explains why small cells, like bacteria, can rely on simple random diffusion to distribute material within them, while larger eukaryotic cells must employ active, directed mechanisms.

\subsubsection{Biomolecules}

	We explore the large-molecule components of cells, DNA, proteins, and lipids, examining especially how their physical attributes are integral to their function. The concept of self-assembly is central, and we examine how the combination of simple physical forces and ubiquitous Brownian motion generates structure. Protein folding provides an important example of this. Building on information about protein sizes and prior exposure to the diffusivity of small molecules, students can estimate the timescale required for a chain of amino acids to explore configuration space and adopt a shape. There are numerous connections between this topic and issues of contemporary interest, even beyond the roles of particular proteins, for example the computational challenges of predicting protein folding outcomes \cite{dill2012}, and the consequences of misfolding in diseases such as mad cow disease and Kuru \cite{pruisner1995}. (The latter, spread by cannibalism, is particularly entertaining to discuss.) Further aspects of protein structure can be explored in group projects, described below. Lipid membranes provide still further opportunity to examine self-assembly, as well as enabling connections to earlier discussions of surface tension and to contemporary research into the mechanical properties of these biological structures \cite{parthasarathycurvature2007, lingwood2010, HonerkampSmith2012}.

DNA is the most iconic biomolecule, and we examine some of the physics related to its role as a conveyor of genetic information. The packaging issues associated with DNA are easy to introduce. We note that each of us have roughly one meter of DNA in each of our roughly one-micron-diameter cell nuclei. We ask: Is this impressive? Since 1 m is much larger than 1 $\mu$m, an obvious answer is ``yes.''  However, we then ask for a simple estimate of the volume of the nucleus, $\sim (10^{-6}m)^3$, and the volume of the DNA, $\sim 1 \mathrm{m} \times (10^{-9}\mathrm{m})^2$, finding that they are similar, so an equally straightforward answer to our question is ``no.''  Both responses are inadequate, however.  To answer meaningfully, we must consider the mechanical properties of DNA. A simple way to do so \cite{pboc} that follows naturally from earlier course topics is to model DNA as a random walk of straight segments, each of length equal to the molecule's persistence length, $\approx 50$ nm. The characteristic size of such a walk, or equivalently the size of a ``blob'' of DNA on its own, is about 200 $\mu$m, showing that its packaging inside the nucleus is, indeed, impressive.

Physics highlights the remarkable challenges involved in packing DNA, and also illuminates the tactics employed in response: DNA is highly negatively charged, and positively charged histone complexes serve as spools on which DNA is wound. The topic of DNA  again connects with issues of scaling, and also links to contemporary studies on, for example, the even denser packing of DNA in many viruses \cite{evilevitch2003}, and the feedback between DNA packaging and the genetic code \cite{segal2006}. Moreover, it highlights the process of model construction in science, and leads to discussions of the motivations, the limitations, and the utility of models.

\subsection {Other topics}

The themes of the course offer abundant opportunities for extensions and personalization of topics, incorporating modern insights into entropy in biomolecular systems, pattern formation, energy flows, experimental tools, and countless other topics.

For example, in some terms we have explored the fundamentally different ways in which large and small organisms must propel themselves in fluids,  \cite{vogelfluids, purcell1977}, a topic that transcends the microscopic and macroscopic divide. This again serves to illustrate that many living creatures inhabit a strange and alien world, at odds with the intuition we develop as large animals in turbulent surroundings.

\section{Components of the Course}

The non-standard subject matter of the course and its audience of non-science-major undergraduates, who are in general rather averse to mathematics, present challenges for teaching that I have attempted to address through the development of a variety of course materials and activities.

	Most class sessions involve a small amount of time spent lecturing, with the considerable majority of the period devoted to active learning in one of several forms. In general, the benefits of active learning methods on student performance are increasingly well appreciated for introductory courses for science majors \cite{deslauriers2012, smith2009, freeman2014}. For general education courses this is has been much less explored, but I would argue that active learning is even more important in this context. Engagement with the material is critical, and since the students in general are less interested in science than are science majors, reliance on passive absorption of new concepts and techniques is not very effective. In addition, students are often trained by prior experiences to believe that they are incapable of scientific inquiry. Making a large fraction of the course require the construction of questions, discussion with peers, and other activities not only helps address this, but does so in a way that makes it clear that this way of learning is the expected norm for the course. 

The most significant tools to aid active learning that we have employed are ``clicker''-based questions about scientific concepts, graphs, or in-class demonstrations\cite{beatty2006}, and in-class worksheets. I have found the worksheets to be highly effective. In these, particular lessons are broken down into a series of questions and discussion topics that students work on in small groups, while teaching assistants and I walk through the class offering advice and asking questions. After most groups have answered a few of the questions, or if many groups are stuck, we all reconvene to go over the topic. For example, our worksheet on the physics underlying bone dimensions began with an exercise plotting bone length and diameter on a logarithmic graph, asked students to sketch graphs that would correspond to isometric- and mechanically-similar scaling, continued with further questions, and finally concluded with a question that encapsulates these concepts and leads to discussion: ``Why can't elephants jump?'' (A few sample worksheets are provided as supplementary materials to this paper \cite{worksheets}.)

There is no textbook that spans the range of subjects described above. Assigned readings of excerpts from Steven Vogel's excellent books on biomechanics \cite{lifesdevices, vogelcats, vogelfluids, vogelglimpses, vogelcircuits}, especially \textit{Life's Devices}\cite{lifesdevices}, and McMahon and Bonner's \textit{On Size and Life} \cite{mcmahon1983}, are useful for the macroscopic half of the term. (The two named books also inspired the creation of the course.) John Tyler Bonner's \textit{Why Size Matters: From Bacteria to Blue Whales}\cite{bonnersizematters} is also elegant and clear, and D'Arcy Thompson's classic \textit{On Growth and Form}\cite{OnGrowthandForm} contains innumerable inspiring morphological discussions. I have supplemented excerpts from books with short articles from \textit{Scientific American}\cite{basu2007, clements1962}, \textit{Physics Today}\cite{west2004}, and other sources, as well as materials I wrote myself. The microscopic half of the course relies much more on readings I have written, and brief excerpts from the publicly available \textit{Molecular Biology of the Cell} textbook \cite{mboc}. A list of assigned  readings is supplied as Supplementary Material \cite{readinglist}. More than half of the class sessions had an assigned prior reading, with a short quiz on its contents at the start of the period.

In addition to readings directly related to class topics, students were given three assignments, to be completed in small groups, in which they read and responded to ``popular science'' articles from \textit{The New York Times}, \textit{The Economist}, and other sources. These dealt with subjects that intersected with those covered in class, for example on scaling relationships claimed to be obeyed by cities, the creation of synthetic nucleotides for DNA, etc. For each article chosen, students were directed to briefly summarize the article, especially the scientific motivations of the work described; to ask one ``quantitative'' question that was not presented in the article or suggest something that could be graphed that would be insightful; and to comment on relationships between the article and in-class topics. Especially since the students in the class are non-science majors, their interaction with science in the future is likely to be largely via popular media of various sorts, and so developing practice with thoughtfully examining popular articles is valuable.

	We also use more standard assignments and assessments: weekly homework assignments and exams. These focus especially on conceptual understanding of the material and order-of-magnitude numerical estimates.

	The course has also incorporated a final project in which students, in groups, research some protein with the goal of explaining the relationship between its structure and function. In addition, students also 3D printed a physical model of their protein and compared the usefulness of this visualization with computational rendering (using the widely used PyMol software). (Interestingly, nearly everyone preferred, overall, the computational illustration.) One could easily imagine a greater focus in the course on scientific visualization methods. 

\section{Challenges and Outcomes}

	``The Physics of Life'' has been offered at the University of Oregon four times  since 2011, each time with an enrollment of about 60. As intended, students represented many different majors (45), and the majority (78\%) were not science majors. There has been a roughly 2:1 ratio of social science to humanities majors. Of students in the natural sciences, the largest contingent (48\%) were psychology majors, and 10\% (3\% of the total students) were physics majors. The course, therefore, succeeded in its aim of reaching a large number of students not pursuing degrees in the sciences.

	Student reaction to the course and subject matter has been enthusiastic.  In written end-of-term comments, many students have noted the ``incredibly interesting and diverse'' topics, and have ``enjoyed how the material we were learning about related to our everyday lives.'' The University of Oregon Science Literacy Program has been surveying attitudes toward and perceptions of science among students in this and other courses; the results will be documented in the near future \cite{slppaper}. Course evaluations, which tabulate responses to a standard set of university-wide questions, do not provide a meaningful measure of student learning, but are unfortunately the only tool available for inter-course comparisons. Evaluation scores for ``The Physics of Life'' are slightly higher than average for general-education courses offered by the University of Oregon Physics Department (``The Physics of Light and Color,'' ``The Physics of Sound and Music,''  and several others). In the evaluation category of overall course quality, for example, the most recent evaluation score was 4.1 out of 5.0, with the departmental mean and standard deviation for the last 76 general-education courses taught being 3.9 $\pm$ 0.3. 
	
Though the course is in general well liked, it presents challenges for students. Many find the mathematical concepts introduced in it difficult. Though nominally simpler than the basic skills in algebra they all have, techniques such as numerical estimation, adeptness with exponents, etc., move beyond their prior habituation with rote memorization of formulae, and require developing a deeper understanding of quantitative perspectives that is non-trivial. For example: via discussion as well as a ``diagnostic'' math quiz during the first week of the term, it is apparent that the considerable majority of students will correctly respond with $x^{a+b}$ when asked what $x^a x^b$ is. However, a much smaller fraction ($\sim 25 \%$), when asked a question like ``if $y$ is proportional to $x^3,$ and $x$ doubles, what happens to $y$?,'' will answer correctly. The former question involves, in most students' experience,  memorization of a rule about manipulating symbols; the latter involves an understanding of what exponents mean. Addressing this, while rewarding and ultimately satisfying, takes time. We do a variety of exercises that build from seemingly trivial beginnings, tabulating the volumes and surfaces areas of simple geometric shapes and plotting them versus linear dimensions on logarithmic axes, and measuring the volumes and masses of isometric (same-shape) objects like bolts, establishing how to think about non-linear relationships, and realizing for example that volume is more than the outcome of formulas about shape, but rather that property of space that scales as length cubed.
	
	The tangibility of the topics explored in the course, i.e. their applicability to the everyday world of animals and plants, helps students engage with physical concepts. Moreover, connections between the course and physiological relevance are particularly valuable, helping to stimulate interest and appreciation. It is not uncommon for students to have  personal experience, for example via family members, with diseases that connect to biophysical properties. (Pre-term births, cystic fibrosis, and cancer have all come up in the course.) Along similar lines, the course serves to illustrate the importance of non-genetic ``information'' in orchestrating life, a message of particular contemporary importance given the tendency of popular media to convey the impression that genes are the sole drivers of function, that there is a gene ``for'' every attribute of health or disease. There is, students learn, no gene that directs lipids into a bilayer, or that ferries neurotransmitters across a chemical synapse; in these and countless other cases, the physical forces and interactions of biomolecules govern and constrain their behavior, a perspective that is important to convey.

	I will also note that the course is very enjoyable to teach. Being a biophysicist, I am, of course, biased, but having taught various other general education courses in recent years, it is apparent that the variety of the subject matter, its connections to the living world around us, and the contemporary excitement of the field of biophysics all make a general education biophysics course a deeply satisfying and intellectually exciting vehicle with which to convey the message of scientific literacy to a general population.

\begin{acknowledgments}

The development and implementation of this course have benefitted enormously from its affiliation with the University of Oregon Science Literacy Program (SLP) \cite{slp}, launched with a grant from the Howard Hughes Medical Institute (grant no. 52006956, Science Education Division) and also supported by funds from the University of Oregon. 

I also gratefully acknowledge input and insights from Elly Vandegrift (associate director of the SLP), Julie Mueller (Univ. of Oregon Teaching Effectiveness Program), Professor Eric Corwin (who taught one term of Physics 171), graduate student assistants Liesl Van Ryswyk, Matt Jemielita, Tristan Hormel, Ryan Baker, Kyle Lynch-Klarup, Savannah Logan, and undergraduate student assistants Kendra Nyberg and Ricky Holton. Undergraduate assistants were upper-division science major supported by the SLP to develop their teaching and communication skills, and to help implement active-learning activities in class. Many the graduate student assistants were wholly or partially supported by SLP.

\end{acknowledgments}

\end{document}